\documentclass[sigconf]{acmart}

\usepackage{comment}
\usepackage{graphicx}
\usepackage{subcaption}
\usepackage{listings}
\usepackage[frozencache,cachedir=.]{minted}
\usepackage{tabularx}

\usepackage{hyperref}
\usepackage{url}
\usepackage{tabularx} %

\usepackage{xcolor}

\definecolor{codegreen}{rgb}{0,0.6,0}
\definecolor{codegray}{rgb}{0.5,0.5,0.5}
\definecolor{codepurple}{rgb}{0.58,0,0.82}
\definecolor{backcolour}{rgb}{0.95,0.95,0.92}

\lstdefinestyle{mystyle}{
    backgroundcolor=\color{backcolour},   
    commentstyle=\color{codegreen},
    keywordstyle=\color{magenta},
    numberstyle=\tiny\color{codegray},
    stringstyle=\color{codepurple},
    basicstyle=\ttfamily\footnotesize,
    breakatwhitespace=false,         
    breaklines=true,                 
    captionpos=b,                    
    keepspaces=true,                 
    numbers=none,                    
    numbersep=5pt,                  
    showspaces=false,                
    showstringspaces=false,
    showtabs=false,                  
    tabsize=2
}

\copyrightyear{2024}
\acmYear{2024}
\setcopyright{rightsretained}
\acmConference[CODS-COMAD Dec '24]{8th International Conference on Data Science and Management of Data (12th ACM IKDD CODS and 30th COMAD)}{December 18--21, 2024}{Jodhpur, India}
\acmBooktitle{8th International Conference on Data Science and Management of Data (12th ACM IKDD CODS and 30th COMAD) (CODS-COMAD Dec '24), December 18--21, 2024, Jodhpur, India}
\acmPrice{}
\acmDOI{10.1145/3703323.3703742}
\acmISBN{979-8-4007-1124-4/24/12}

\begin{document}

\title{Static Program Analysis Guided LLM Based Unit Test Generation}

\author{Sujoy Roychowdhury}
\authornote{Equal Contribution}
\affiliation{%
  \institution{Ericsson R\&D}
  \city{Bangalore}
  \country{India}
}

\author{Giriprasad Sridhara}
\authornotemark[1]
\affiliation{%
  \institution{Ericsson R\&D}
  \city{Bangalore}
  \country{India}
}

\author{A K Raghavan}
\authornote{The author was at Ericsson R\&D during this study}
\affiliation{%
  \institution{Independent Researcher}
  \city{Chennai}
  \country{India}}

\author{Joy Bose}
\affiliation{%
  \institution{Ericsson R\&D}
  \city{Bangalore}
  \country{India}
}

\author{Sourav Mazumdar}
\affiliation{%
  \institution{Ericsson R\&D}
  \city{Bangalore}
  \country{India}
}

\author{Hamender Singh}
\affiliation{%
  \institution{Ericsson R\&D}
  \city{Bangalore}
  \country{India}
}

\author{Srinivasan Bajji Sugumaran}
\affiliation{%
  \institution{Ericsson R\&D}
  \city{Bangalore}
  \country{India}
}

\author{Ricardo Britto}
\affiliation{%
  \institution{Ericsson}
  \city{Stockholm}
  \country{Sweden}
}

\renewcommand{\shortauthors}{Roychowdhury et al.}

\begin{abstract}
We describe a novel approach to automating unit test generation for Java methods using large language models (LLMs). 
Existing LLM-based approaches rely on sample usage(s) of the method to test (focal method) and/or provide the entire class of the focal method as input prompt and context. The former approach is often not viable due to the lack of sample usages, especially for newly written focal methods. The latter approach does not scale well enough; the bigger the complexity of the focal method and larger associated class, the harder it is to produce adequate test code (due to factors such as exceeding the prompt and context lengths of the underlying LLM).
We show that augmenting prompts with \emph{concise} and \emph{precise} context information obtained by program analysis 
increases the effectiveness of generating unit test code through LLMs.
We validate our approach on a large commercial Java project and a popular open-source Java project.
\end{abstract}

\ccsdesc[500]{Computing methodologies~Natural language generation}
\ccsdesc[300]{Computing methodologies~Discourse, dialogue and pragmatics}
\ccsdesc[500]{Software and its engineering}

\keywords{Unit Test Case Generation, Large Language Models, Static Analysis}
\begin{teaserfigure}
\centering
  \includegraphics[width=0.7\textwidth]{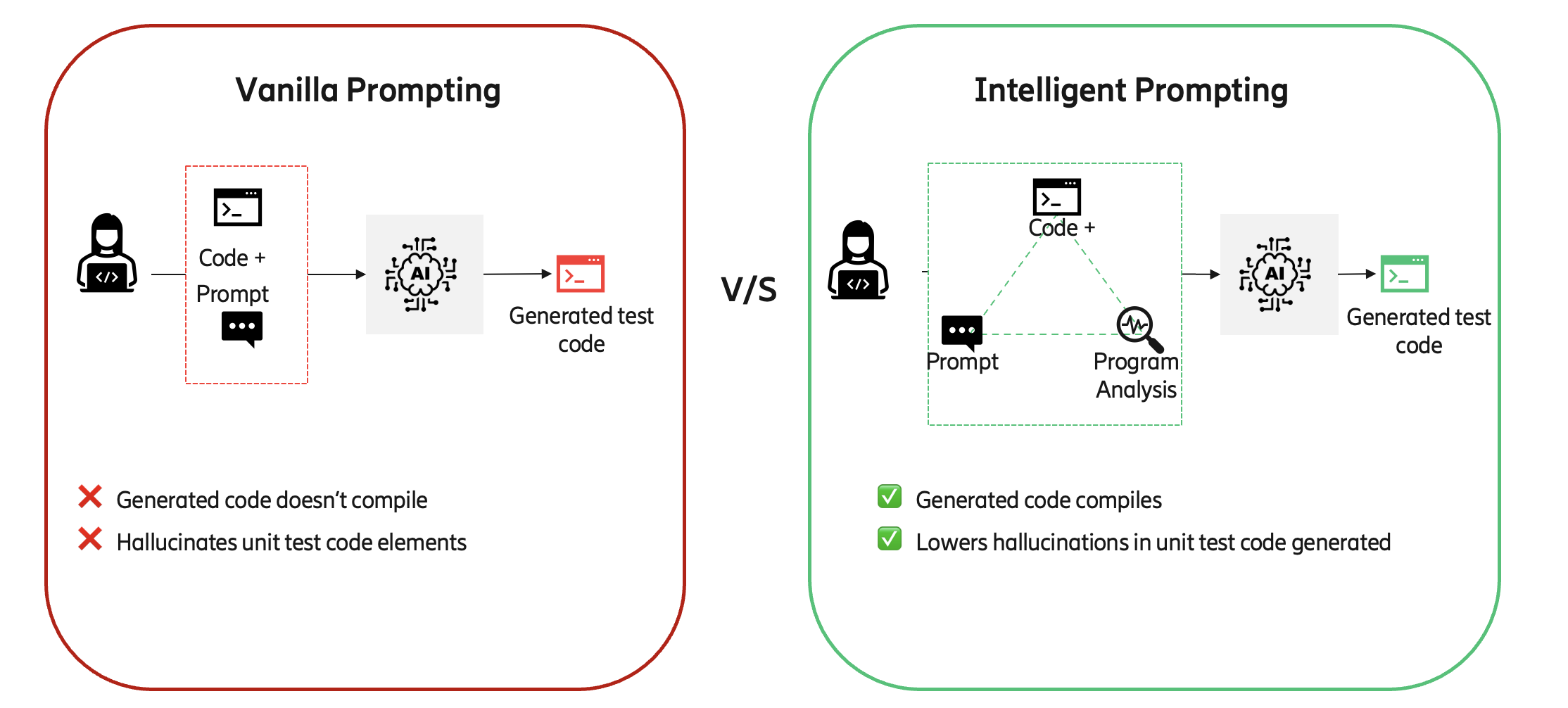}
  \caption{Unit Test Case Generation enhanced with Program Analysis}
  \label{fig:teaser}
\end{teaserfigure}


\maketitle

\section{Introduction}\label{sec:Introduction}

\begin{listing*}[h!]
\begin{lstlisting}[
language=Java,
frame=lines,
framesep=1mm,
basicstyle=\normalsize,
numbers=left
]
String getEmail() { return isEmailUnknown() ? getUnknownEmail(number) : email; }

Simple Prompt: Please generate Unit Test
  LLM Generated Unit Test (Only Line of Interest shown):
    when(myClass.getUnknownEmail(anyString())).thenReturn("Unknown");
         
Program Analysis Enhanced Prompt: Please generate Unit Test. getUnknownEmail() defined in class 
'Employee'. Signature of getUnknownEmail = String getUnknownEmail(Integer)

  LLM Generated Unit Test (Only Line of Interest shown):
    when(employee.getUnknownEmail(anyInt())).thenReturn("Unknown");
\end{lstlisting}
\caption{Motivating Example: Line 1 is the Focal Method. Line 5 has two compilation errors which are remedied in Line 11 due to the program analysis information added to the prompt.}
\label{motiv-ex}
\end{listing*}

Unit tests are an essential part of software quality assurance~\cite{daka2014survey}. 
They are complete functions (methods) with one or more statements that invoke the \emph{method under test} (\emph{also, Focal Method}). They have at least one assertion or verification statement about the result accruing from calling the \emph{Focal Method}. 
The advent of LLMs\cite{openai2023gpt4,rozière2023code,touvron2023llama} has enabled researchers to explore their use for unit test  generation~\cite{utg-llm-usage-example-based,utg-llm-full-file-context}. However, these approaches have 
limitations that prevent their usage for unit test generation in large commercial software. This paper discusses these shortcomings and proposes a novel approach to address them.

The approach described in ~\cite{utg-llm-usage-example-based} relies on \emph{sample usages} of the \emph{focal method} (i.e., method under test).
In our use case, especially for newly written focal methods, such example usages are not readily available. Thus, we cannot use this approach.

An orthogonal approach to unit test generation using LLMs treats the problem as a code completion task~\cite{utg-llm-full-file-context}. It  provides the entire source file in which the focal method is defined and the initial parts of an unit test file with comments describing that the test file contains unit tests.
We use this approach as a \emph{baseline} in our evaluation, and show that this approach leads to a large prompt and context and \emph{unfortunately} it does not generate unit tests for many focal methods in our code base.

We alleviate the above problem by adroitly ensuring that the prompt and the context is as \emph{precise} and \emph{concise} as possible using static program analysis which we describe in Section ~\ref{sec:approach}.We augment the prompt with information obtained from static program analysis of the focal method. Our technique ensures that only relevant information is provided to the LLM and hence the context length is not exceeded enabling the LLMs to generate syntactically correct unit tests.

The main contributions of this paper are as follows:
\begin{enumerate}
    \item A novel and efficient approach for unit test generation using LLMs whose prompts are enhanced \emph{precisely} and \emph{concisely} with static program analysis; and, 
    \item An evaluation of the above approach on a commercial closed source Java Project and an open source Java project and comparing against the baseline approach~\cite{utg-llm-full-file-context}. 
\end{enumerate}

The remainder of this paper is organized as follows. Section~\ref{sec:RelatedWork} describes the related work while Section~\ref{sec:staticProgramAnalysis} provides a high level background about the static program analysis used in this paper. Section~\ref{sec:Methodology} describes our methodology and Section~\ref{sec:eval} depicts our evaluation. Section~\ref{sec:Results} delineates our results. In Section~\ref{sec:discussion} we provide a discussion of the results. We conclude in Section~\ref{sec:conc} along with a discussion of future work.

\section{Related Work}\label{sec:RelatedWork}

RLPG (Repo Level Prompt Generator) \cite{shrivastava2023repository} describes an approach to single-line code auto completion task by providing additional contextual information such as information about other project files, in the prompt. We differ in that our task is \emph {an entire method generation} i.e., unit test generation. 
Importantly, we provide a restricted set of additional contextual information due to the character limits on the prompt and input to the LLM. 
Monitor Guided Decoding \cite{agrawal2023monitor} is a white box approach to LLMs that interferes at the output generation part of the LLM. In contrast, we do \emph{not need} such an access to the internal weights of the LLM.

There has been a large body of work in automatically generating unit tests. These include non-Artificial Intelligence techniques such Evosuite~\cite{evosuite} and Randoop~\cite{randoop}.
Evosuite implements a search-based generation technique, while, Randoop implements a feedback-directed random technique.

Then we have unit test generation techniques based on machine and deep learning. Here unit test generation is treated in a manner similar to a natural language translation task (say English to French). The method under test (focal method) is provided as input, with its matching unit test as the output. Large number of such pairs are used in an encoder-decoder network to learn and generate unit tests for hitherto unseen focal methods~\cite{tufano2020unit}.

Unit tests need an oracle that can specify the correct or the expected output from it. Automatically generating such oracles using neural networks has been explored in~\cite{dinella2022toga}. Once an oracle is known for a unit test, we need \emph{assertions} that compare the expected and the actual output in different ways. Automatically generating such \emph{assert} statements using transformers has been described in~\cite{tufano2022generating}.

\section{Background - Static Program Analysis}\label{sec:staticProgramAnalysis}

Static Program Analysis deals with analyzing the source code of a program without executing the program and observing the results (which is called dynamic program analysis). 
Static Program Analysis is heavily used in Software Engineering to detect problems such as null pointer exceptions; to automatically complete source code and so on.

Static Program Analysis involves several parts such as parsing the code, building Control Flow Graphs to explore control flow within the program, data flow analysis to detect data flow and so on.

Automated Java Parsers operate using the Backus Naur Form (BNF) Grammar of the Java Language and detect program entities such \textcolor{black}{as} method (function) declarations, method invocations (function calls). Parsers construct an Abstract Syntax Tree from the program source code text. Symbolic Solvers analyze the AST and informs about the types of the program entities (For example, in Listing~\ref{motiv-ex}, the getUnknownEmail method requires one parameter whose data type is Integer ).

\section{Methodology: Program Analysis Guided Unit Test Generation via LLMs}\label{sec:Methodology}

\label{sec:approach}

Our approach along with the motivating example is shown in Listing~\ref{motiv-ex}. The method under test is shown on Line 1. Line 3 shows a simple prompt and Line 5 shows one line of interest in the LLM generated unit test in response to the prompt. Lines 7 and 8 depict \emph{another} prompt augmented with information obtained by a static program analysis of the code (including the method under test and the project). Line 11 is the counterpart of line 5 and shows the new code generated.

Juxtaposing lines 5 and 11, we see that line 5 has two compilation errors, viz., the use of the \emph{receiver object} i.e., \emph{myClass} and the assumption about the parameter of \emph{getUnknownEmail} being a \emph{String}. In Line 11, due to the additional program analysis information, the LLM uses a correct object of the \emph{Employee} class and correctly uses \emph{anyInt()} to denote an \emph{Integer} parameter.


We perform automated static program analysis using the Java Parser Library. For each focal method, we obtain the declaring class of the method, the complete signatures of the set of methods called from the focal method and the data type information about the fields used in the focal method. We then prompt the LLM to generate unit test(s) using the JUnit framework. We also prompt the LLM to use \emph{Mocking}, which is a way to isolate the dependencies of the focal method and concentrate on its behaviour.

It is especially important for mocking that the method signatures be known precisely in a strongly typed language like Java. Thus, this is where our addition of the program analysis information particularly helps.

\section{Evaluation}\label{sec:eval}

\textbf{Experiment Subjects:} We evaluated our work on a large commercial Java Project from a telecom company. 
We also used the popular open source Java project, Guava.

\textbf{LLMs:} We used llama {\textcolor{black}{7b/70b 2.0}} and CodeLlama {\textcolor{black}{34b based on llama 2.0}} as the LLMs in our experiments as these have a commercial friendly license and have been shown to be amongst the better peforming LLMs. {\color{black} In addition, we have extended analysis on open-source code to GPT-4 for benchmarking. We are not allowed to use propietary code on GPT-4.} 

\textbf{Evaluation Metric:} Unit tests can be evaluated using a variety of factors such as syntactic and semantic correctness; assertions used, test smells and so on. However, before using these facets, test cases should actually be generated. We found that the baseline could not generate test cases for many focal methods. Thus, in this paper, we compare the number of focal methods for which one or more unit tests were generated by the baseline and our approach. We asked and answered the following research question:

\textbf{RQ1:} Does our approach of adding precise and concise program structural information to the prompt increase the number of focal methods for which one or more unit tests are generated?

\subsection{Dataset and Experiments}\label{subsec:dataset}
We selected a random sub-project from our commercial Java project and this had 103 focal methods.
We then performed the following two experiments: 1) A baseline approach as described in~\cite{utg-llm-full-file-context};and,
(2) Our approach where for each focal method we provided only the focal method's signature and body along with  the program structural information as shown in Listing ~\ref{our-prompt} in Appendix.

For the open source project, Guava, we selected two random Java classes and repeated the above experiments with the 34 public methods in these classes.

\subsection{Implementation Details}\label{subsec:Implementation}

\subsubsection{LLM Parameters}\label{subsubsec:LLMParms}

For inference using the LLMs we keep the input tokens limited to 1023 (the llama APIs mandate that it cannot be larger) and the full context length is default of 4096. Temperature is set to zero, topK is set to 50 and topP is set to 0.95. Other parameters are kept at their default. The models are running on a single A100 80 GB GPU and we use 8 bit  quantization for inference.

\subsubsection{Experiment Setup}\label{subsec:experimentSetup}
We use the Java Parser library APIs to automatically parse the project Java code. For each method under test, we obtain the complete signatures of the invoked methods, such as, \emph{int java.util.List getSize()}. We also obtain the declared type of any field accessed within the method. We augment the prompt with these additional program context information and provide the focal method to the LLM and collect the output i.e., the generated unit tests. We used llama and code llama as the LLMs as we needed an LLM that could be hosted locally and which had a commercial friendly license (Apache).

\section{Results}\label{sec:Results}

\begin{table*}[h!]
\centering
\small
\begin{minipage}{0.48\textwidth}
    \centering
    \begin{tabular}{|c|c|c|c|} \hline
         & \textbf{llama7b} & \textbf{llama70b} & \textbf{codellama34b} \\ \hline
      Baseline &  37 & 9 & 7\\
       Ours  &  \textbf{102} & \textbf{88} & \textbf{63} \\ \hline
    \end{tabular}
    \caption{103 focal methods from commercial: For how many were unit tests generated?}
    \label{tab:tests-gen-commercial}
\end{minipage}
\hfill
\begin{minipage}{0.48\textwidth}
    \centering
\begin{tabular}{|c|c|c|c|c|} 
\hline
 & \textbf{llama7b} & \textbf{llama70b} & \textbf{codellama34b} & \textbf{\color{black}{gpt-4}} \\ 
\hline
Baseline &  20 & 20 & 1 & \textcolor{black}{32} \\ 
Ours  &  \textbf{30} & \textbf{31} & \textbf{19} & \textcolor{black}{\textbf{34}} \\ 
\hline
\end{tabular}

    \caption{34 focal methods from open source: For how many were unit tests generated?}
    \label{tab:tests-gen-guava}
\end{minipage}
\end{table*}

\begin{table*}[h!]
\centering
\small
\begin{minipage}{0.48\textwidth}
    \centering
    \begin{tabular}{|c|c|c|c|} \hline
         & Mean & Std. Dev. & Median \\ \hline
      Baseline   &  5295 & 3350 & 4056\\
       {\color{black}{Ours}}  &  559 & 516 & 464 \\ \hline
    \end{tabular}
    \caption{Tokens in input} 
    \label{tab:tokens}
\end{minipage}
\hfill
\begin{minipage}{0.48\textwidth}
    \centering
    \begin{tabular}{|c|c|c|c|} \hline
         & Mean & Std. Dev. & Median \\ \hline
      Baseline   &  38 & 32 & 22\\
        {\color{black}{Ours}} &  32 & 13 & 30 \\ \hline
    \end{tabular}
    \caption{Time Taken in Seconds}
    \label{tab:time}
\end{minipage}
\end{table*}


\subsection{Prompts and Sample Outputs}\label{subsec:sampleOutputs}

We present in this section a sample prompt for baseline and ours and output from the LLM on the prompts.

\textbf{Our Prompt:}
Listing ~\ref{our-prompt} shows a sample of the prompt used in our approach. It is for the method \textit{ensureNonNegative} in the class \emph{DoubleUtils} from Google's popular open source Java project, \emph{Guava}. This is one of the two random classes we used from \emph{Guava} in our open source evaluation.

\begin{lstlisting}[caption={Sample of our prompt: Method from Google Guava Project},label={our-prompt}]
[INST]
<<SYS>>
Generate Java unit test(s) for the given Java focal method. Mock method calls and fields using Mockito library. Use the JUnit framework and try to obtain high branch coverage.
<</SYS>>
FOCAL-METHOD-BEGIN
static double ensureNonNegative(double value) {
    checkArgument(!isNaN(value));
    return Math.max(value, 0.0);
}
FOCAL-METHOD-END
Declaring-Class-of-Method:
DoubleUtils
SIGNATURES-OF-METHOD-CALLS-WITHIN-FOCAL-METHOD:
void com.google.common.base.Preconditions.checkArgument(boolean)
double java.lang.Math.max(double, double)
boolean java.lang.Double.isNaN(double)
FIELDS-USED-WITHIN-FOCAL-METHOD:
[/INST]
\end{lstlisting}

\begin{lstlisting}[caption={Output from the LLM for our prompt: One Unit Test Generated with correct assertions and suitably named identifiers and appropriate imports.},label={our-prompt-output}]
import static org.junit.Assert.*;
import org.junit.Test;
import org.mockito.Mockito;
public class DoubleUtilsTest {
    
    @Test
    public void testEnsureNonNegative() {
        // Test with positive input
        assertEquals(1.5, DoubleUtils.ensureNonNegative(1.5), 0.0001);
        
        // Test with negative input
        assertEquals(0.0, DoubleUtils.ensureNonNegative(-2.3), 0.0001);
        
        // Test with NaN input
        assertTrue(Double.isNaN(DoubleUtils.ensureNonNegative(Double.NaN)));
    }
}
\end{lstlisting}

\textbf{Baseline Prompt:} The baseline prompt for the same method, \emph{ensureNonNegative} in the class \emph{DoubleUtils} (guava project) is much longer than our prompt. It has the whole code prompted. \footnote{The relevant code \texttt{DoubleUtils.java} is available at \href{https://github.com/google/guava/blob/master/guava/src/com/google/common/math/DoubleUtils.java}{DoubleUtils.java}}

\textbf{Output from baseline prompt:}
The baseline prompt \emph{did not} generate \emph{any} unit tests. It simply regurgitated the input as the response.

\subsection{Evaluation Results}\label{subsec:EvalResults}

Results are shown in Tables~\ref{tab:tests-gen-commercial} and ~\ref{tab:tests-gen-guava}, for the commercial and open source projects respectively. It shows that across the different LLMs, our approach, is able to generate tests for more focal methods as compared with the baseline.  


\

\section{Discussion}\label{sec:discussion}

Our hypothesis is that the baseline is not able to generate \emph{any} test cases for \emph{many} focal methods due to confounding factors such as increased context length, presenting extraneous information to the LLM. In contrast, our approach presents only \emph{precise} and \emph{concise} information to the LLM and thus can get the LLMs to be successful. To understand this phenomenon better, we obtained statistics on the input token lengths. The baseline has a mean and a median of 5295 and 4056 tokens, respectively. In contrast, the token length in our approach is noticeably smaller, with a mean and median 559 and 464 tokens, respectively. The statistics for tokens and time are presented in Table~\ref{tab:tokens} and Table~\ref{tab:time} respectively.

\paragraph{\textbf{Cost Factor:}} Many LLMs such as the OpenAI based ones have a token based charge system. In such cases, our approach would be further beneficial.



\paragraph{\textbf{Portability:}}\label{subsec:portability}

Although we have demonstrated our approach on Java, it should be portable across languages, as we only need program analysis information which can be obtained automatically for every programming language out there using parsers and symbolic resolvers. The approach should also work for different complex methods as we have chosen methods randomly from open source and closed source (commercial projects).

\section{Conclusion and Future Work}\label{sec:conc}

We described a novel approach to unit test generation using LLMs. The prompt of the LLM is guided by precise and concise information obtained automatically by program analysis. We evaluated our work on a commercial and an open source Java project. Our evaluation shows that our approach is able to generate unit tests for more focal methods as compared with the baseline approach, which presumably fails to due to the complexity and length of its input prompt and context. In future, we will deploy our solution in production and perform additional experiments using other metrics on commercial and open source software {\color{black} as well as extend to other programming languages}. {\color{black} One limitation of this study is that we are only looking at generated test cases rather than their quality - this is a scope for future study. Also not all dependencies, especially those at runtime, can be analyzed via static program analysis - such limitations would affect our results too - however quantifying the impact of the same may be challenging. Further work involves extending our method to an In-context learning (ICL) or fine tuning of LLMs.}

\bibliographystyle{ACM-Reference-Format}
\bibliography{COMADS_TestCaseGen_for_Arxiv/testCaseGen}


\end{document}